\documentclass[aps,prl,twocolumn,showpacs]{revtex4}
\usepackage{graphicx}

\begin{document}

\title{Phase and vortex correlations in Josephson-junction arrays
at irrational frustration}

\author{Enzo Granato}

\address{Laborat\'orio Associado de Sensores e Materiais,
Instituto Nacional de Pesquisas Espaciais,  12245-970 S\~ao Jos\'e
dos Campos, SP Brazil}

\begin{abstract}
Phase coherence and vortex order in a Josephson-junction array at
irrational frustration are studied by extensive Monte Carlo
simulations using the parallel tempering method. A scaling analysis
of the correlation length of phase variables in the full
equilibrated system shows that the critical temperature vanishes
with a power-law divergent correlation length and critical exponent
$\nu_{ph}$, in agreement with recent results from resistivity
scaling analysis. A similar scaling analysis for vortex variables
reveals a different critical exponent $\nu_{v}$, suggesting that
there are two distinct correlation lengths associated with a
decoupled zero-temperature phase transition.

\end{abstract}


\maketitle

Josephson-junction arrays at irrational frustration have attracted
considerable interest, both experimentally and theoretically, as a
possible physical realization of a two-dimensional vortex glass or a
pinned incommensurate vortex lattice, without intrinsic disorder.
Frustration without disorder can in principle be introduced by
applying an external magnetic field in a perfect periodic array of
weakly coupled superconducting grains \cite{carini,zant,baek} and
similarly in superconducting wire networks \cite{yu,ling}. The
frustration parameter $f$, the number of flux quanta per plaquette,
sets the average density of the vortex lattice and can be tuned by
varying the strength of the external field \cite{teitel}. At
rational $f$, the ground state is a pinned vortex lattice
commensurate with the array leading to discrete symmetries in
addition to the continuous $U(1)$ symmetry of the phase variables
characterizing the superconducting order parameter. The phase
transitions and resistive behavior of the array are only reasonably
well understood for simple rational $f$. At irrational $f$, however,
when the vortex-lattice is incommensurate with the array, both the
nature of the equilibrium phase transition and of the
low-temperature state in the thermodynamic limit remain unclear.

We consider a Josephson-junction array on a square lattice described
by the Hamiltonian  \cite{teitel}
\begin{equation}
H=-J\sum_{<ij>}\cos(\theta_i -\theta_j-A_{ij}) .  \label{model}
\end{equation}
$\theta_i$ is the phase of the local superconducting order
parameter, $J>0$ is the uniform Josephson-junction coupling and
$A_{ij}$ is constrained to be $\sum_{ij}A_{ij} = 2 \pi f$ around
each plaquette, where $f$ is an irrational number
$f=(3-\sqrt{5})/2$, related to the Golden Ratio
$\Phi=(1+\sqrt{5})/2$ as $f=1-1/\Phi$.

In early Monte Carlo (MC) simulations \cite{halsey}, the ground
state was found to consist of a disordered vortex pattern and a
possible superconducting (vortex-glass) transition at a finite
temperature $T_g \sim 0.25$ was proposed from the behavior of the
specific heat. On the other hand, some arguments suggested that the
critical temperature $T_c$ should vanish \cite{teitel,choi}.
Simulations of the current-voltage scaling found indeed a behavior
consistent with a $T_c=0$ phase transition \cite{eg96} similar to
the vortex glass model in two dimensions \cite{fisher,eg98}, but
with a different correlation-length critical exponent $\nu \sim 1$.
However, since resistivity scaling probes mainly phase coherence,
the behavior of the vortex correlation still remained to be
investigated. Simulations of the relaxation dynamics by Kim and Lee
\cite{kimlee} of the vortex variables found a behavior analogous to
supercooled liquids with a dynamic crossover temperature close to
the apparent  $T_g$ observed earlier in the specific heat
\cite{halsey}. MC simulations \cite{teitelf}, using the vortex
representation for rational $f$ converging to the irrational
frustration, suggested two phase transitions at finite temperatures,
a first-order transition to an ordered vortex structure weakly
dependent on $f$ and a phase-coherence transition at much lower
temperatures varying significantly with $f$.  The results were in
qualitative agreement with other MC simulations using the phase
representation \cite{tang}, but different ground states were found
and the first-order behavior was also sensitive to the boundary
conditions.

More recently, a study of the finite-size behavior of the specific
heat and relaxation time in the phase representation found an
intrinsic finite-size effect \cite{park}. The scaling analysis
confirmed the $T_c=0$ transition scenario with $\nu$ consistent with
the earlier estimate from current-voltage scaling \cite{eg96}. Other
simulations agree that below some temperature relaxation processes
become very slow. Very recently, the $T_c=0$ scenario received
further support from improved calculations using a driven MC
dynamics \cite{eg07}. On the other hand, an analysis of the
low-temperature configurations for $f$ close to the irrational value
from  MC simulations in the vortex representation \cite{llkim}
suggested two transitions, consistent with earlier work
\cite{teitelf}.

In view of these conflicting results, it is important to determine
the true equilibrium behavior using  methods that insure full
equilibration of the system and obtain the critical behavior
directly from the phase and vortex-correlation lengths within the
same framework.

In this work we study phase coherence and vortex order at irrational
frustration by extensive MC simulations, using the
parallel-tempering method (exchange MC method) \cite{nemoto} to
obtain equilibrium configurations of the system. This method has
been shown to reduce significantly the long equilibration times in
glassy systems \cite{nemoto,cooper,ly} and supercooled liquids
\cite{kolb}. To study the equilibrium phase transitions we use
numerical data in the temperature regime in which full equilibration
can be insured and employ a scaling analysis to extrapolate to the
low-temperature and large-system limits. Since finite-size scaling
of the correlation length is currently one of the most reliable
approaches to demonstrate the existence of an equilibrium
finite-temperature transition for glassy systems \cite{cooper,ly},
we use this  analysis for the phase and vortex variables. The
results indicate that the critical temperature for phase coherence
vanishes ($T_c=0$) with a power-law divergent correlation length and
corresponding critical exponent $\nu_{ph}$, in agreement with recent
results from resistivity scaling \cite{eg07}. Although a first-order
vortex transition at finite temperatures can not be ruled out, a
similar scaling analysis for vortex variables is also consistent
with $T_c=0$  but with a different exponent $\nu_{v}$. These
different exponents suggest the interesting scenario where there are
two distinct correlation lengths associated with a decoupled $T_c=0$
phase transition.

In the numerical simulations we use periodic boundary conditions on
lattices of linear sizes $L$ and corresponding rational
approximations $\Phi_n=F_{n+1}/F_n$, where $F_n$ are Fibonacci
numbers ($5,8,13,21,34,55$), with $L=F_n$. Additional calculations
using periodic (fluctuating twist) boundary conditions \cite{eg07}
or the exact value of $f$, did not change the results.

To study phase coherence we consider the overlap order parameter
\cite{bhatt} of the phase variables defined as
$q_{ph}(j)=\exp(i\theta_j^1-i\theta_j^2)$, where 1 and 2 denote two
thermally independent copies of the system with the same parameters
$J$ and $f$. At high temperatures, where each copy is thermally
disordered, the correlation function
$C_{ph}(r)=\frac{1}{L^2}\sum_j<q_{ph}(j)q_{ph}(j+r)>$ is short
ranged, decaying exponentially with $r$, while at low temperatures
it is long ranged if an  ordered phase exists, including the
possibility of a glassy-ordered phase. The corresponding correlation
length in the finite-size system $\xi_{ph}$ can be obtained from a
second moment calculation using the correlation function as
\cite{cooper}
\begin{equation}
\xi_{ph}(L)=\frac{1}{2\sin(k_o/2)}(\frac{S_{ph}(0)}{S_{ph}(k_o)}-1)^{1/2}
,
\end{equation}
where $S_{ph}(k)$ is the Fourier transform of $C_{ph}(r)$  and
$k_o=(\frac{2\pi}{L},0)$ is the smallest wave vector in the finite
system. The same expressions  are used to determine the correlation
length for vortex variables $\xi_{v}$  in terms of the vorticity
$v_p$, replacing $q_{ph}$ by $q_v(p)=v_p^1 v_p^2$. The vorticity is
defined as  $v_p=\sum_{ij}(\theta_i-\theta_j-A_{ij})/2\pi$ and is  a
measure of the local vortex density, where the summation is taken
over the elementary plaquette $p$ of the lattice and the
gauge-invariant phase difference is restricted to the interval
$[-\pi,\pi]$.

\begin{figure}
\includegraphics[bb= 2cm 3.5cm  19cm   11cm, width=7.5 cm]{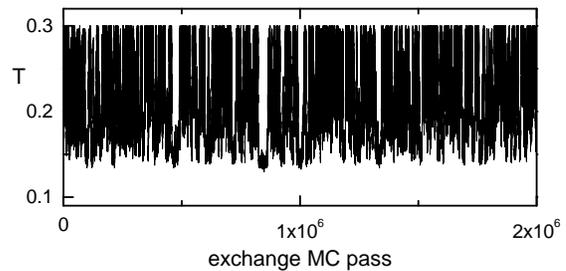}
\caption{ Trajectory in temperature space of a replica starting at
temperature $T=0.30$, for system size $L=55$. The simulation
included $100$ replicas in the range $T=0.30$ to $T=0.096$ .}
\end{figure}

We use the parallel tempering method \cite{nemoto} to obtain the
equilibrium configurations. Many replicas of the system with
different temperatures are simulated simultaneously and the
corresponding configurations are allowed to be exchanged with a
probability satisfying detailed balance. The exchange process allows
the configurations of the system to explore the temperature space,
being cooled down and warmed up, and the system can escape more
easily from metastable minima at low temperatures. With this method,
full equilibration can be insured in finite size systems
\cite{nemoto,ly,kolb}. Without the replica exchange step, the method
reduces to conventional MC simulations at different temperatures. We
performed MC simulations using the heat-bath algorithm for each
replica, simultaneously and independently, for a few MC passes. Then
exchange of pairs of replica configurations at temperatures $T_i$
and $T_j$ and energies $E_i$ and $E_j$ is attempted with probability
$min(1,\exp(-\Delta))$, where $\Delta = (1/T_i - 1/T_j)(E_j -E_i)$,
using the Metropolis scheme. The equilibration time to reach thermal
equilibrium can be measured as the average number of MC passes
required for each replica to travel over the whole temperature
range. We used typically $4 \times 10^6$ MC passes for equilibration
with up to $100$ replicas and equal number of MC passes for
calculations of average quantities. Nevertheless, for the largest
system sizes $L=21-55$, equilibration was only possible for
temperatures above $T_f \sim 0.145$. This can be inferred from the
time evolution, in the temperature space, of a replica initially at
the highest temperature, as shown in Fig. 1. The replica
configuration starting at  $T = 0.3$ is only able to explore the
temperature space containing $100$ temperatures down to $T_f \sim
0.145$. Below this temperature, the replicas cannot be warmed up and
cooled down. Thus $T_f$ can be regarded as a freezing temperature,
below which the system remains trapped in metastable configurations
within the available time scale of the present simulation. In fact,
below $T_f$ the numerical results for $\xi_{ph}$ and $\xi_{v}$ are
sensitive to the initial conditions while above $T_f$ they are not.
Our estimate of $T_f$ is well below the apparent glass temperature
$T_g$, observed in earlier MC simulations \cite{halsey}.

For the finite-size scaling analysis of the correlation
length \cite{cooper,ly}, we consider the dimensionless ratio $\xi/L$
which, for a continuous transition, should satisfy the scaling form
\begin{equation}
\xi/L=G((T-T_c)L^{1/\nu}) \label{scalxi}
\end{equation}
where $\nu$ is the critical exponent of the power-law divergent
correlation length $\xi\propto |T-T_c|^{-\nu}$, $T_c$ is the
critical temperature and $G(x)$ is a scaling function with $G(0)=C$,
a constant, and $G(x)\rightarrow x^{-\nu}$ as $x\rightarrow \infty
$. This scaling form implies that data for the scaled correlation
length $\xi/L$ as a function of temperature, for  different system
sizes $L$, should come together for decreasing temperatures and
cross at the same temperature $T=T_c$. In addition, the data should
splay out for different system sizes with slopes determined by the
critical exponent $\nu$.

\begin{figure}
\includegraphics[bb= 2cm 4cm  19cm   16cm, width=7.5cm]{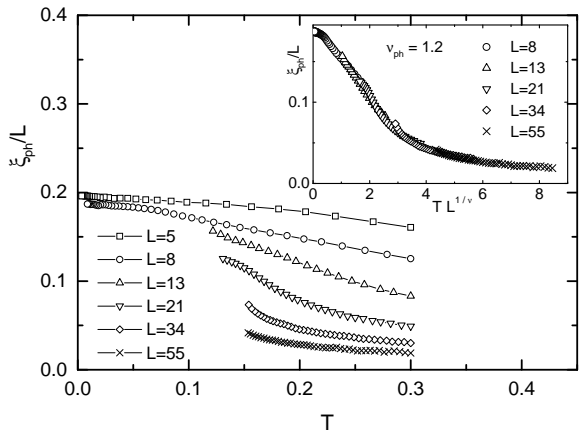}
\caption{ Scaled correlation length of phase variables $\xi_{ph}/L$
for different system sizes $L$. Inset: scaling plot according to Eq.
\ref{scalxi}, assuming $T_c=0$ with $\nu_{ph}=1.2$. }
\end{figure}

Fig. 2 shows the temperature dependence of the scaled correlation
length for phase variables, $\xi_{ph}/L$, in the temperature range
where full equilibration was possible and for different system
sizes. This quantity increases faster on lowering the temperature as
the system size $L$ increases indicating a divergent length scale
for decreasing temperature. However, for fixed temperature it
decreases with $L$ even at the lowest available temperature and
therefore the curves do not cross at a common temperature. If a
phase-coherence transition takes place then it should occur at some
unknown critical temperature $T_c$ much below $T_f \sim 0.145$,
which is not accessible in our calculations for larger system sizes,
or else only at $T=0$. The latter case corresponds to a transition
where $T_c=0$ and the correlation length $\xi_{ph}$ is finite at any
nonzero $T$  but diverges when approaching $T=0$. In principle,
requiring that the data should satisfy the scaling form of Eq.
\ref{scalxi}, could be used to determine $T_c$ and consequently find
out which scenario is realized. However, for $T_c
> 0$ such data collapse needs two different adjustable parameters,
$T_c$ and $\nu$, which it is not a sufficiently accurate method. On
the other hand, the $T_c=0$ scenario can be verified more accurately
since the scaling analysis requires adjusting only the critical
exponent $\nu$. In this case, the data for $\xi_{ph}$ should satisfy
the finite-size scaling form of Eq. \ref{scalxi} with $T_c=0$ and
the best data collapse provides an estimate of the critical exponent
$\nu_{ph}$. Fig. 2 (inset) shows that indeed the data satisfy the
scaling form with an exponent $\nu_{ph}=1.2(2)$.

\begin{figure}
\includegraphics[bb= 2cm 3.5cm  19cm   13cm, width=7.5 cm]{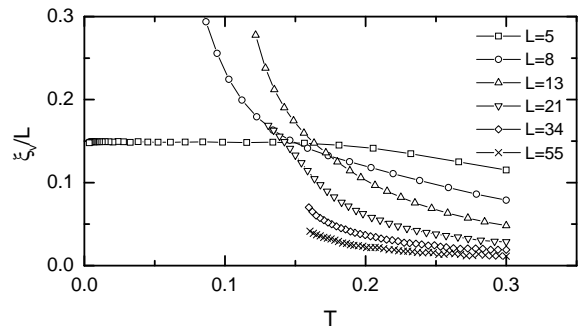}
\caption{ Scaled correlation length of vortex variables $\xi_{v}/L$
for different system sizes $L$. }
\end{figure}

The $T_c=0$ scaling behavior for phase coherence and the exponent
$\nu_{ph}$ are in agreement with results obtained from resistivity
scaling using the RSJ model for the dynamics \cite{eg96} and, more
recently, resistivity scaling using a driven MC dynamics
\cite{eg07}. Although $T_c=0$, at finite temperatures the relevant
divergent correlation length determines both the linear and
nonlinear resistivity of the array leading to a current-voltage
behavior described by the scaling theory.  In the present case,
where we can define two correlations lengths, $\xi_{ph}$ and
$\xi_v$, the relevant divergent quantity should be $\xi_{ph}$ since
this is  a measure of phase coherence. From the resistivity scaling
the estimate was \cite{eg07} $ \nu_{ph}= 1.4(2)$, which agrees
within the errors with the present direct estimate from correlation
length calculations. This quantitative agreement for the value of
$\nu_{ph}$ obtained from  equilibrium and dynamical calculations
provides strong support for the phase-coherence transition scenario
\cite{eg96,eg07} with $T_c=0$.

\begin{figure}
\includegraphics[bb= 2cm 4cm  19cm   13cm, width=7.5 cm]{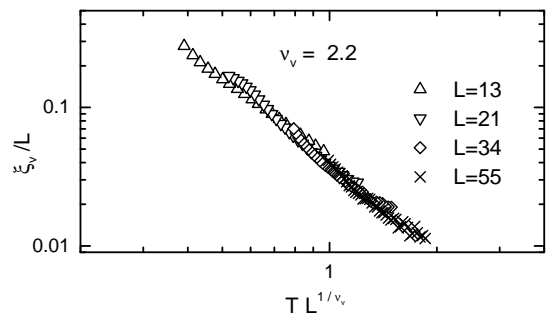}
\caption{ Scaling plot according to Eq. \ref{scalxi}, assuming
$T_c=0$, for the vortex correlation length $\xi_{v}$, with
$\nu_v=2.2$. }
\end{figure}

In Fig. 3 we  show the finite-size behavior of the scaled
correlation length for vortex variables $\xi_{v}/L$. This quantity
also increases faster on lowering the temperature. For small system
sizes ($L=5-13$) the curves intersect at different temperatures near
$T\sim 0.18$ but for larger system sizes they all decrease with $L$
even at the lowest available temperature. Again, the lack of
intersection at a common temperature for large system sizes suggests
that vortex order, or even vortex-glass order, may only occur at
$T_c=0$ or $T_c << T_f$. Alternatively, this lack of intersection at
a common temperature may suggest a vortex first-order transition.
Such a transition was  proposed earlier based on results from MC
simulations in the vortex representation \cite{teitelf}, where also
a phase-coherence transition at much lower temperature was observed.
The first order transition was suggested from the observation of a
double peaked energy distribution near the transition. These results
are in qualitative agreement with MC simulations using the phase
representation \cite{tang}. However, different ground states were
found and the double peaked energy distribution was also sensitive
to the boundary conditions. Moreover, to confirm the first order
nature of the transition, a finite-size scaling analysis of the
energy distribution using much large system sizes would be required.
Since the parallel-tempering method used here is known to be a
significant improvement over conventional MC methods by allowing
escape from metastable configurations and reducing the equilibration
time at low temperatures \cite{nemoto}, the lack of fully
equilibration that we found for $T < T_f$ makes unclear wheather the
double peaked energy distributions observed in finite systems in
other MC simulations \cite{teitelf,tang} are the result of an
underlying equilibrium first order transition in the thermodynamic
limit or of a (non equilibrium) freezing transition. Thus, although
a first-order vortex transition at finite temperature can not be
ruled out, the possibility remains that this transition is second
order and actually occurs at zero temperature. In this case, the
data for $\xi_{v}$ for the largest system sizes should satisfy the
finite-size scaling form of Eq. \ref{scalxi} with $T_c=0$ and the
best data collapse provides an estimate of the critical exponent
$\nu_v$. Fig. 4 shows that indeed the data satisfy this scaling
form. Surprisingly, however, the estimated critical exponent for
vortex variables $\nu_v=2.2(3)$ is significantly different from the
one for phase variables $\nu_{ph}=1.2(2)$. This in turn suggests
that the $T_c=0$ critical behavior is not described by a single
divergent length scale and therefore that there is a decoupling of
phase and vortex correlations both diverging as a power-law as
temperature approaches zero, but with different critical exponents.

It is clear from Figs. 2 and 3 that both correlation lengths
$\xi_{ph}$ and $\xi_v$ remain finite at $T \le T_g \sim 0.25$, the
apparent glass temperature found in earlier MC simulations
\cite{halsey}, since the ratio $\xi/L$ decreases with system size.
Therefore, the signature of glass behavior found in this earlier
work should be attributed to slow dynamics effects and not an
equilibrium phase transition.

In conclusion, our scaling analysis is consistent with a $T_c=0$
transition \cite{teitel,choi,eg96,eg07} but the phase and vortex
correlation lengths diverge with different critical exponents
suggesting a new decoupled zero-temperature transition scenario.

Work supported by FAPESP (Grant 07/08492-9).

\end{document}